# Efficient all-perovskite tandem solar cells by dual-interface optimisation of vacuum-deposited wide-bandgap perovskite

Yu-Hsien Chiang,[a] Kyle Frohna,[a] Hayden Salway,[b] Anna Abfalterer,[a] Bart Roose,[b] Miguel Anaya,[*b] Samuel D. Stranks[*a.b]

a. Cavendish Laboratory, Department of Physics, University of Cambridge, JJ Thomson Avenue, Cambridge CB3 0HE, United Kingdom

b. Department of Chemical Engineering & Biotechnology, University of Cambridge, Philippa Fawcett Drive, Cambridge CB3 0AS, United Kingdom

*sds65@cam.ac.uk, *ma811@cam.ac.uk

# Abstract

Tandem perovskite solar cells beckon as lower cost alternatives to conventional single junction solar cells, with all-perovskite tandem photovoltaic architectures showing power conversion efficiencies up to 26.4%. Solution-processing approaches for the perovskite layers have enabled rapid 2optimization of perovskite solar technologies, but new deposition routes are necessary to enable modularity and scalability, facilitating further efficiency improvements and technology adoption. Here, we utilise a 4-source vacuum deposition method to deposit $FA_{0.7}Cs_{0.3}Pb(I_xBr_{1-x})_3$ perovskite, where the bandgap is widened through fine control over the halide content. We show how the combined use of a MeO-2PACz self-assembled monolayer as hole transporting material and passivation of the perovskite absorber with ethylenediammonium diiodide reduces non-radiative losses, with this dual-interface treatment resulting in efficiencies of 17.8% in solar cells based on vacuum deposited perovskites with bandgap of 1.76 eV. By similarly passivating a narrow bandgap $FA_{0.75}Cs_{0.25}Pb_{0.5}Sn_{0.5}I_3$ perovskite and combining it with sub-cells of evaporated $FA_{0.7}Cs_{0.3}Pb(I_{0.64}Br_{0.36})_3$, we report a 2-terminal all-perovskite tandem solar cell with champion open circuit voltage and power conversion efficiency of 2.06 V and 24.1%, respectively. The implementation of our dry deposition method enables high reproducibility in complex device architectures, opening avenues for modular, scalable multi-junction devices where the substrate choice is unrestricted.

# Introduction

Multi-junction solar cells constitute the most practical way to achieve power conversion efficiencies (PCEs) beyond the radiative efficiency limits of single-junction solar cells. Multi-junction technologies employ photo-absorbers with complementary bandgaps to collectively harvest a broader portion of the solar spectrum whilst minimising thermalisation losses upon hot-carrier relaxation. The highest performance solar cell report to date is from a triple-junction based on III-V semiconductors with a strain-balanced quantum well stack, achieving a PCE of 39.5%.[1] However, their high cost due to complex fabrication processes that involve high temperatures limits their accessibility and versatility. These costs are historically limiting the use of III-V solar cells in terrestrial power applications and restrict their use to high-value applications such as powering satellites or space vehicles.

Halide perovskites are generating enormous excitement as thin film absorbers for high-performance solar cells, showing a unique combination of features that include low-temperature processing and a resilience to electronic defects.[2] With a compositionally tunable $ABX_3$ crystal structure, where A = methylammonium (MA), formamidinium (FA) and/or Cs, B = Pb and/or Sn, and X = Cl, Br, I, the bandgap of 3D perovskites can be varied from 1.2 eV to 3.0 eV. This absorption tunability, combined with high absorption coefficients and charge carrier mobilities, make these materials promising for both single and multi-junction thin film solar cells.[3–6] Indeed, the record PCE single-junction perovskite and all-perovskite tandem solar cells have reached PCEs of 25.7% and 26.4%,[7,8] respectively, representing the most efficient emerging PV systems to date. These outstanding outcomes result from years of work from myriads of research groups mostly working with solution processed approaches that allow rapid screening and optimisation. However, solution approaches ultimately present limitations for manufacturing due to the use of toxic solvents and potential issues with dissolving underlying layers, the latter limiting the underlying materials and substrates.

Vacuum deposition processes show great promise to overcome barriers related to large-area coating, integration into flexible, light-weight substrates and novel device patterns while ensuring high thickness control and conformal film uniformity, all with a solvent-free technique. To date, fully evaporated perovskite solar cells have achieved PCEs of 20.7% and 21.4% on small active area (< 0.2 cm$^2$) by co-evaporation and sequential evaporation, respectively,[9,10] and PCE of 18.1% over larger area (21 cm$^2$).[11] While the community has concentrated most efforts on evaporating MAPbI$_3$ perovskite solar cells,[9,11–13] we and others have demonstrated that MA-free, mixed halide systems are viable candidates to achieve thermally stable perovskite devices.[14–17] Importantly, the dry nature of the technique represents an ideal approach to stack different perovskite films for tandem device architectures on a range of underlying contacts and substrates, an approach that opens avenues for a highly efficient yet low-cost thin film, light-weight perovskite technology.

Nevertheless, only a handful of works have reported fully-evaporated perovskites for their application in multi-junction cells, with most of the examples focusing on deposition processes to combine perovskite and silicon subcells in a tandem fashion.[18–20] As for all-perovskite tandem systems, Ávila *et al.* reported a vacuum deposited MAPbI$_3$-MAPbI$_3$ perovskite solar cell with an outstanding V$_{OC}$ of 2.3 eV and a PCE of 18%, demonstrating the potential of the technique to attain building blocks for tandem devices.[21] However, perovskite bandgaps in this work were not optimised to minimise energy losses while maximising current matching for AM1.5 illumination. Optical modelling suggests that a PCE >35% cell efficiency is potentially achievable in all-perovskite tandems under realistic conditions.[22–24] This value is conditioned by the absorption spectrum of the rear subcell as the narrowest perovskite bandgaps demonstrated so far are in the range between 1.20 eV to 1.30 eV based on alloyed Pb/Sn compositions. The bandgap of the optimum front (wide bandgap) subcell for that constraint is between 1.70 to 1.80 eV, though little further loss is seen when the bandgap is lowered further to 1.65 eV when light coupling between layers is taken into account.[25] One challenge in realising these wide bandgap perovskite materials is they inevitably require mixed halide compositions, and they hence suffer from light-induced phase segregation, forming Br and I rich sub domains[26–29] that reduce the open-circuit voltage ($V_{OC}$).[30] Lidón *et al.* reported a wide bandgap (1.77 eV) perovskite with a composition of $FA_{0.61}Cs_{0.39}Pb(I_{0.70}Br_{0.30})_3$ displaying a $V_{OC}$ of up to 1.21 V, the best-to-date for a vacuum deposited system.[14] Yet, this $V_{OC}$ is still 240 mV below the radiative efficiency limit of 1.45 V for a 1.77 eV bandgap, indicating there are still significant losses in the best wide bandgap perovskite solar cells. Interestingly, it has been recently shown that low radiative efficiency in bulk mixed halide perovskites and energy misalignment between the perovskite and contact layers are the main losses in wide bandgap perovskite solar cells, and a device $V_{OC}$ of over 1.33 V (for a 1.77 eV bandgap) is achievable even in the presence of halide segregation.[30,31] These results overall show the complex compromise between perovskite phase stabilisation and device stack optimisation required to attain vacuum deposited wide bandgap perovskite solar cells relevant for tandem architectures. In this work, we employ a 4-source co-evaporation technique to systematically vary the bandgap of $FA_{0.7}Cs_{0.3}Pb(Br_xI_{1-x})_3$ films from 1.62 eV to 1.80 eV for their subsequent integration in an all-perovskite tandem device. We show how contact layer optimisation by using (2-(3,6-Dimethoxy-9H-carbazol-9-yl)ethyl)phosphonic acid (MeO-2PACz) as hole transporting material (HTM) minimises interfacial recombination and leads to PCEs of 20.7% for a 1.62 eV perovskite, which is among the highest PCE reported for a vacuum deposited perovskite system. The addition of higher Br fractions to blueshift the absorption onset for wide bandgap subcells introduces defects, as demonstrated by a reduction in the photoluminescence quantum efficiency (PLQE) and a higher Urbach energy, which is particularly exacerbated at 1.80 eV where substantial phase segregation readily occurs. We demonstrate that ethylenediammouiunm diiodide (EDAI) is an effective passivation agent for these evaporated perovskites, resulting in PLQEs enhanced by an order of magnitude. Applying the HTM and EDAI dual-interface treatment yields devices with a $V_{OC}$ of 1.26 V for a 1.76 eV bandgap, which is 190 mV below the radiative limit and represents the lowest $V_{OC}$ loss reported in evaporated wide bandgap perovskite systems so far. Furthermore, we demonstrate the versatility of the EDAI passivation approach by applying it to solution-processed, narrow bandgap perovskite solar cells displaying a $V_{OC}$ of 0.86 V based on MA-free, Pb/Sn absorbers with a bandgap of 1.28 eV. An EDAI-passivated 2-terminal tandem architecture combining the evaporated $FA_{0.7}Cs_{0.3}Pb(I_{0.64}Br_{0.36})_3$ perovskite with the solution-processed $FA_{0.75}Cs_{0.25}Pb_{0.5}Sn_{0.5}I_3$ perovskite results in a PCE of 24.1%, the highest for evaporation-based all-perovskite tandem solar cells. This result shows the potential of the scalable and industry-relevant evaporation technique for realising efficient and modular all-perovskite tandem solar cells.

## Hole transporting/perovskite interface optimisation

Defects at the perovskite-hole transporting layer interface are known to cause significant non-radiative losses in p-i-n devices,[33] limiting their applicability for tandem architectures where high voltages are required. Al-Ashouri *et al* reported that the non-radiative losses arising from the interface between perovskite and the typically employed poly[bis(4-phenyl)(2,4,6-trimethylphenyl)amine] (PTAA) can be substantially reduced when replacing the latter by a self-assembled monolayer (SAM), MeO-2PAC or 2-PACz, leading to higher device $V_{OC}$.[25,26] In order to explore this effect and optimize MA-free evaporated systems, we fabricate devices with an architecture consisting of ITO / HTM / perovskite (500 nm) / C60 (25 nm) / BCP (8 nm) / Cu. We initially employ our recently reported 3-source evaporation protocol[17] to deposit $FA_{0.7}Cs_{0.3}Pb(I_{0.9}Br_{0.1})_3$ absorbers (Fig. 1a with no PbBr$_2$) on different HTMs, namely 2-PACz, PTAA and MeO-2PAC, with the absorber exhibiting a bandgap of 1.62 eV extracted via the inflection point of an external quantum efficiency (EQE) measurement (Supplementary Fig. 1). Scanning electron microscopy (SEM, Supplementary Fig. 2) images do not show significant differences in surface morphology, suggesting that the perovskite growth is similar on these different organic layers We observe reduced non-radiative recombination in the perovskite/MeO-2PACz structure, with the PLQE a factor of 3.3 and 5 times higher than perovskite deposited on 2-PACz and PTAA, respectively (Supplementary Fig. 3). Figure 1b shows that the MeO-2PACz based devices display a substantially higher $V_{OC}$ (1.11 V on average) and less $V_{OC}$ variation between devices than the other HTMs, with the trend consistent with higher PLQE[34] and across different batches (Supplementary Fig. 4). Indeed, the devices based on 2-PACz show s-kinks in the current-voltage (J-V) curves, resulting in very low performance (Supplementary Fig. 5 and Table S1). This result is in quite stark contrast to the high performance achieved in solution-processed systems on 2-PACz, even though we employ the same deposition parameters for all HTMs.[34,35] We note that re-optimisation of different evaporated perovskite compositions and deposition parameters might yield improved performance on 2PACz. Our champion device reaches a PCE of 20.7% when the evaporated perovskite is deposited on MeO-2PACz (Figure 1c), which is the highest MA-free perovskite solar cells reported for multisource evaporation. Further, the device retains 90% of its initial performance after 120 hours under one sun illumination and a fixed bias at maximum power point with no substrate temperature control during the measurement (Figure 1d). We find that this front interface optimisation is critical to ensure maximised voltages for subsequent integration of the cells as building blocks in tandem devices.

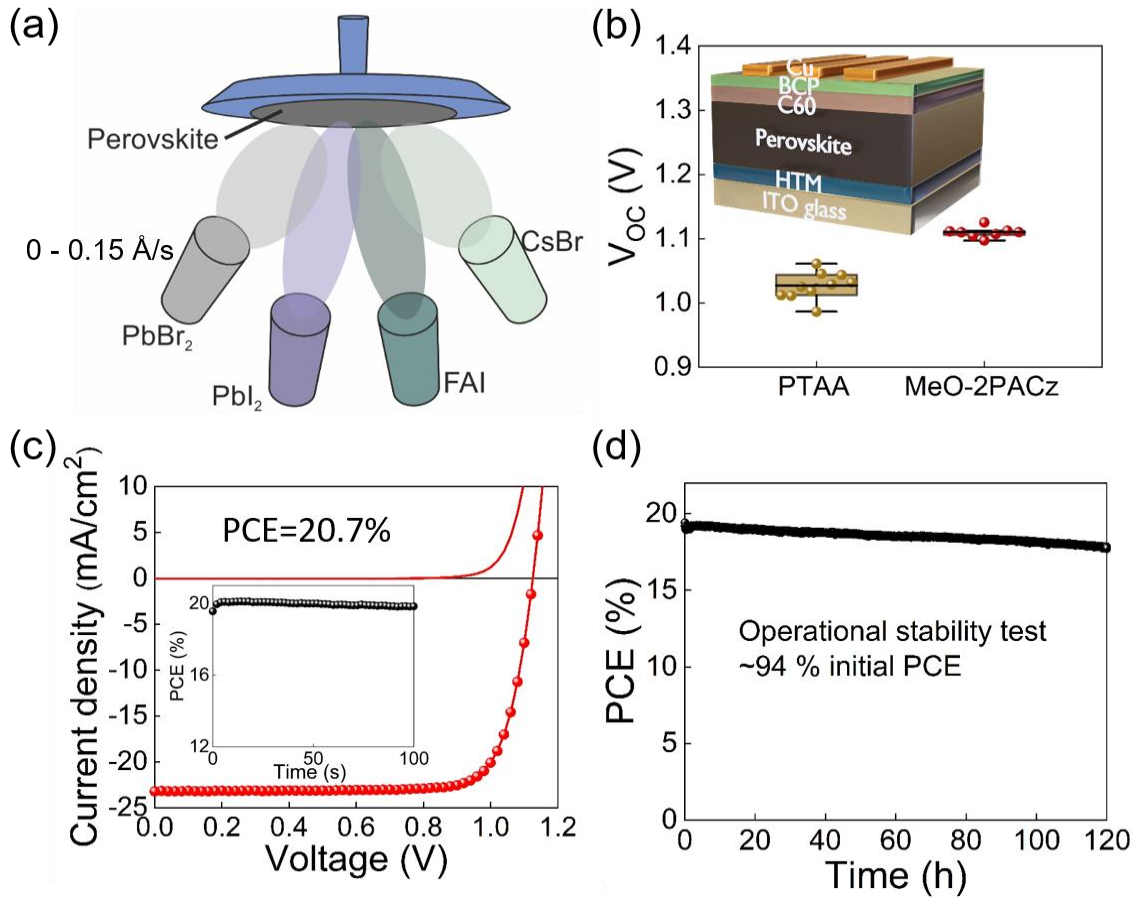

Figure 1. (a) Diagram of the thermal evaporation system where four sources (i.e. FAI, PbBr$_2$, PbI$_2$ and CsBr) are employed to deposit high quality FA$_{0.7}$Cs$_{0.3}$Pb(I$_x$Br$_{1-x}$)$_3$ thin films. FAI, PbI$_2$ and CsBr evaporation rates are kept at 1, 0.6 and 0.1 Å/s, respectively, and the PbBr$_2$:PbI$_2$ deposition rate ratio varied to tune the bandgap. (b) Effect of different HTMs (i.e. PTAA, and MeO-2PACz) on the V$_{OC}$ statistics of evaporated FA$_{0.7}$Cs$_{0.3}$Pb(I$_{0.9}$Br$_{0.1}$)$_3$ perovskite solar cells (PbBr$_2$ rate is 0). Inset device architecture employed. The box/whisker plot contains the 1.5 interquartile range, the median value and data distribution from 8 devices. (c) J-V curve of the champion FA$_{0.7}$Cs$_{0.3}$Pb(I$_{0.9}$Br$_{0.1}$)$_3$ perovskite solar cells with the inset showing a stabilised power output measurement. (d) Operational stability test of an encapsulated device at 0.94 V fixed bias under continuous 1 sun illumination.

## Wide bandgap perovskite deposited via 4-source evaporation

To widen the bandgap of the perovskite for its use as a front subcell in a tandem device, we add PbBr2 as a fourth evaporation source to tune the bandgap by employing PbBr2:PbI2 rate ratios from 0.11 to 0.318. XRD patterns shown in Figure 2a confirm Br incorporation into the perovskite structure as the PbBr2 evaporation rate is increased, with the (011, cubic) perovskite peak shifting to higher 2θ as a consequence of a smaller d-spacing (see Supplementary Fig. 6 for full XRD patterns). Top view SEM images (Supplementary Fig. 7) show perovskite grain sizes in the range of 100 to 300 nm for all compositions as well as the presence of PbI$_2$ evidenced by bright clusters, consistent with the PbI$_2$ signal observed in the XRD patterns.[36] In a previous report, we showed that the presence of excess PbI$_2$ enhances the optoelectronic properties of these vacuum deposited perovskites and their stability when exposed to ambient conditions.[17] We estimate the stoichiometry of the evaporated perovskite films using XRD across a range of control films to generate a calibration curve (Supplementary Fig. 8),[17] and display the resulting chemical formulae in Table 1. We determine the corresponding bandgaps as the inflection point of the first derivative in the EQE spectrum (Supplementary Fig. 1), observing that the bandgap varies between 1.62 to 1.80 eV (Table 1) confirming the increase in bandgap upon additional Br incorporation. To understand this observation and give further insight into the chemical composition, we perform synchrotron-based nano-X-ray fluorescence (nXRF) measurements on our samples. The nanoprobe nature of the technique allows us to extract Br:Pb maps with a spatial resolution of ~50 nm (Figure 2b). Evaporated perovskites with bandgaps between 1.62 eV and 1.77 eV show excellent halide spatial homogeneity, which is particularly striking when comparing to a standard solution-processed 'triple-cation' FA$_{0.79}$MA$_{0.16}$Cs$_{0.05}$Pb(I$_{0.83}$Br$_{0.17}$)$_3$ perovskite film (bandgap of 1.62 eV), where we have found that the compositional heterogeneity is related to defects and carrier funnelling[2,37]. Nevertheless, the 1.80 eV-bandgap evaporated film exhibits several areas with Br-rich clusters, suggesting a suboptimal intermixing of compounds in samples with the highest explored Br content, which is known to drastically hamper stability.[38]

Steady-state photoluminescence (PL) measurements also reflect this variation (Figure 2c), with a clear tunability in the PL peak position from 1.62 to 1.80 eV. Evaluation of the charge carrier lifetime by time-resolved photoluminescence (TRPL) indicates lower optoelectronic quality in the wide bandgap perovskites when compared with the 1.62 eV counterparts (Figure 2d). Both samples show a common quick decay in the first 40 ns attributed to quenching by the contacts. The subsequent PL decay of the 1.77 eV evaporated perovskite deposited on top of the MeO-2PACz/ITO contact shows faster (84 ns) mono-exponential decay with respect to that of the control 1.62 eV sample (394 ns),[39,40] where mono-exponential decays are expected in experiments with such low carrier densities.[41] We attribute the faster decay associated with Shockley–Read–Hall (SRH) recombination to an increase in the trap density when we replace fractions of I by Br in the perovskite composition. We note that the carrier densities may differ with different quenching efficiencies at the contacts between the samples and this may in turn influence the subsequent lifetimes.

A major issue hindering the applicability of wide bandgap perovskites is their phase instability under illumination. To evaluate this, we use a 520-nm continuous-wave laser at 5 suns intensity (300 W/cm$^2$) to excite encapsulated samples and monitor their PL over time. Samples with bandgaps in the range between 1.62 and 1.77 eV show excellent emission stability, with no changes in their PL spectra over time at these photon doses (Figure 2c and Supplementary Fig. 9). In contrast, severe phase segregation occurs in the 1.80 eV perovskite during the first minutes under illumination, potentially linked to the substantial halide heterogeneity observed by nXRF mapping (see Figure 2c).

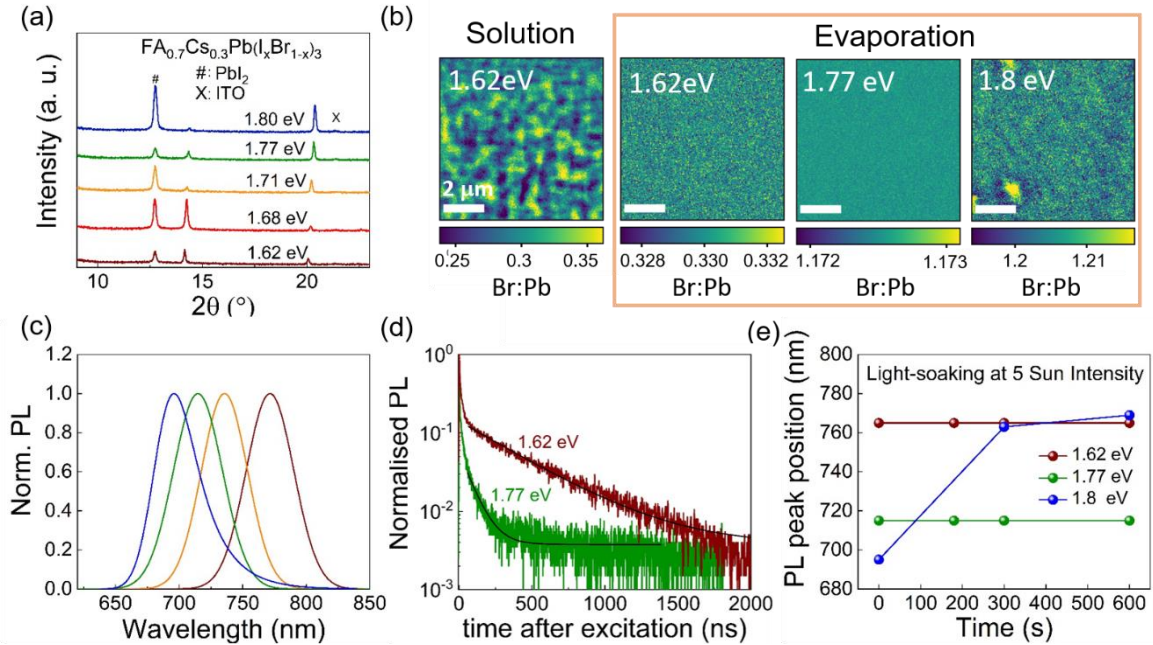

Figure 2. (a) XRD pattern of evaporated $FA_{0.7}Cs_{0.3}Pb(I_xBr_{1-x})_3$ thin films deposited on MeO-2PACz/ITO showing different bandgaps: 1.62 eV (dark red curve), 1.68 eV (red), 1.71 eV (orange), 1.77 eV (green) and 1.80 eV (blue). (b) Br:Pb obtained by nXRF mapping for solution-processed triple-cation perovskite and evaporated perovskites with bandgaps of 1.62 eV, 1.77 eV and 1.80 eV. Scale bars are of 2 μm. (c) PL spectra of a series of evaporated $FA_{0.7}Cs_{0.3}Pb(I_xBr_{1-x})_3$ thin films deposited on MeO-2PACz/ITO showing different bandgaps: 1.62 eV (dark red curve), 1.71 eV (orange), 1.77 eV (green) and 1.80 eV (blue). (d) Time-resolved photoluminescence decays for 1.62 eV and 1.77 eV evaporated perovskite films on MeO-2PACz excited with a 450 nm laser at a fluence of 8.5 nJ/cm$^2$/pulse. Samples were excited and light collected from the top side. We attribute the fast decay during the first 40 ns results from the ITO/ MeO-2PACz quenching. (e) PL peak evolution over time for the 1.62 eV, 1.77 eV and 1.8 eV evaporated perovskite films under continuous illumination at 5 Suns (300 W/cm$^2$) with a 520 nm laser.

Table 1. Champion PV performance metrics for evaporated perovskite solar cells of different bandgap.

| Bandgap (eV) | $V_{OC}$ (V) | $J_{SC}$ (mA/cm$^2$) | FF (%) | PCE (%) | Composition | PbBr$_2$ rate (Å/s) |
|---|---|---|---|---|---|---|
| 1.62 | 1.11 | -23.0 | 78.70 | 20.0 | FA$_{0.7}$Cs$_{0.3}$Pb(I$_{0.9}$Br$_{0.1}$)$_3$ | 0 |
| 1.68 | 1.14 | -19.8 | 78.50 | 17.6 | FA$_{0.7}$Cs$_{0.3}$Pb(I$_{0.78}$Br$_{0.21}$)$_3$ | 0.06 |
| 1.71 | 1.18 | -19.1 | 78.39 | 17.7 | FA$_{0.7}$Cs$_{0.3}$Pb(I$_{0.75}$Br$_{0.24}$)$_3$ | 0.1 |
| 1.77 | 1.24 | -18.5 | 69.22 | 15.9 | FA$_{0.7}$Cs$_{0.3}$Pb(I$_{0.64}$Br$_{0.36}$)$_3$ | 0.127 |
| 1.80 | 1.23 | -17.4 | 72.92 | 15.6 | FA$_{0.7}$Cs$_{0.3}$Pb(I$_{0.56}$Br$_{0.44}$)$_3$ | 0.15 |

We fabricate single-junction solar cells based on the different perovskite compositions to evaluate their performance when integrated into working devices. We use the device architecture introduced in Figure 1a with MeO-2PACz as HTM and show the J-V curves in Figure 3a and EQE spectra in Figure 3b. These measurements demonstrate efficient photocarrier-to-electron conversion for all devices, and a blue shifted absorption onset upon Br addition to the perovskite composition. The device $V_{OC}$ monotonically increases for higher perovskite bandgaps (Figure 3c), with the highest $V_{OC}$ of 1.24 V observed for the 1.77 eV evaporated perovskite (Table 1). There is no further voltage gain for a device based on a 1.80 eV bandgap absorber. We associate this $V_{OC}$ saturation to the substantial phase segregation (cf. Figure 2b,e), which produces low gap clusters onto which charge carriers funnel. To gain further understanding on our device losses, we calculate the estimated $V_{OC,rad}$ based on the Urbach fit of EQE spectra to obtain the extended EQE tail for dark current calculation (Supplementary Fig. 10; see Methods for more details) and represent the $V_{OC}$ loss associated with the different evaporated composition in Figure 3d.[42,43] We observe that the $V_{OC}$ loss increases from 196 mV to 251 mV when we tune the perovskite bandgap from 1.62 eV to 1.80 eV. The Urbach energy also rises with the bandgap energy from 13.5 meV to 19.0 meV. These observations indicate higher electronic disorder upon Br addition and are consistent with the increased trap densities revealed from the PL measurements (cf. Fig. 2d).[2] These collective performance and photo-stability results suggest that the evaporated $FA_{0.7}Cs_{0.3}Pb(I_{0.64}Br_{0.36})_3$ perovskite with a 1.77 eV bandgap is our best candidate for use as front absorber in a tandem architecture with sufficient photo-stability and remarkable, albeit still suboptimal $V_{OC}$.

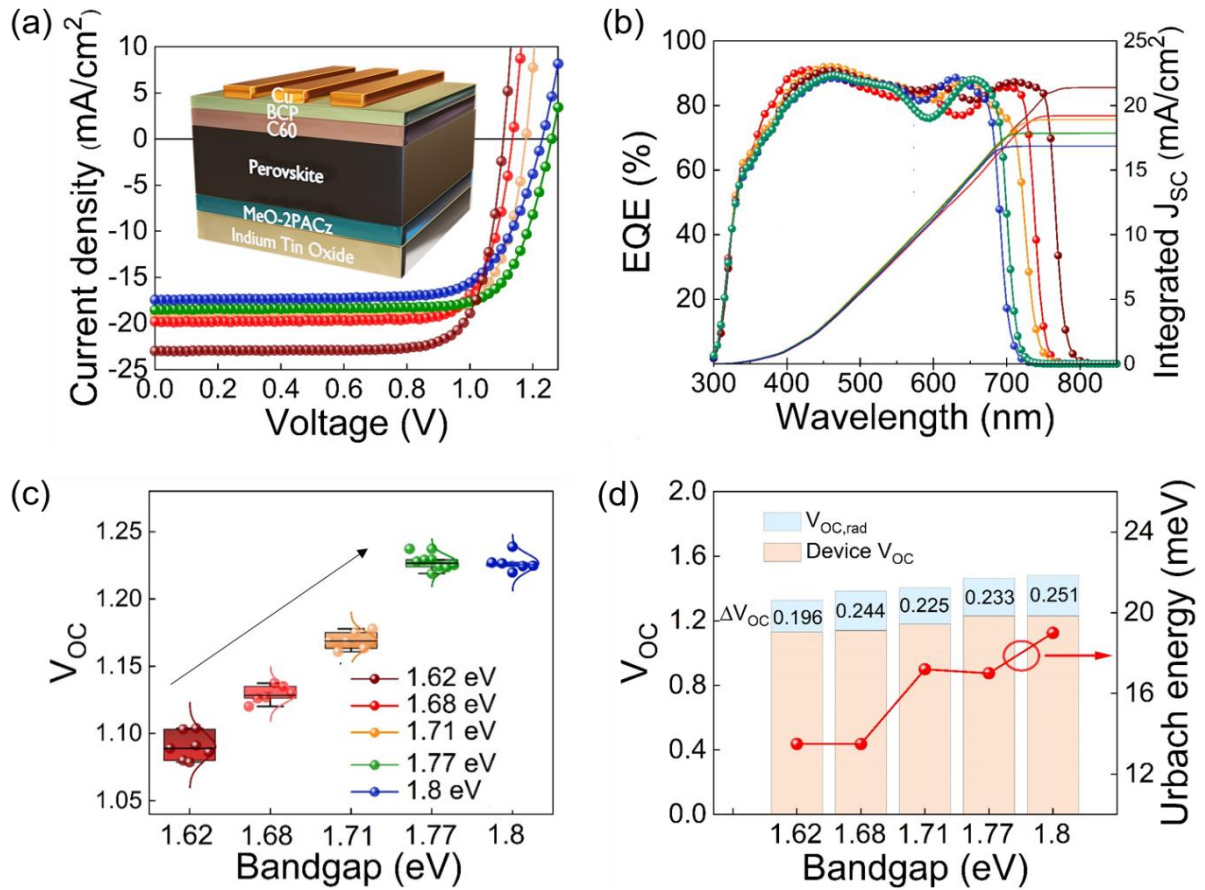

Figure 3. (a) J-V curves at AM1.5G illumination of evaporated solar cells based on perovskites with different bandgaps with colour legend as in panel c: 1.62 eV (dark red curve), 1.68 eV (red), 1.71 eV (orange), 1.77 eV (green) and 1.80 eV (blue). See Table 1 for compositions. EQE spectra and the corresponding integrated $J_{SC}$ (b) and the $V_{OC}$ statistics (c) for the series of devices shown in panel (a). Box/whisker plots contain the 1.5 interquartile range, the median value, and data distribution of the $V_{OC}$. (d) Measured $V_{OC}$, calculated $V_{OC}$ loss and Urbach energy extracted from EQE for champion devices based on different bandgap, evaporated perovskite.

**Perovskite/electron transporting layer interface optimisation**

With the losses associated with the front perovskite interface minimised by employing an HTM based on MeO-2PACz, we now focus on overcoming the losses arising from the rear perovskite interface. To this end, we passivate the 1.77 eV evaporated $FA_{0.7}Cs_{0.3}Pb(I_{0.64}Br_{0.36})_3$ perovskite using a post-treatment by spin-coating a layer of EDAI in mixed isopropanol and toluene (see Methods).[32] We observe an order-of-magnitude improvement in PLQE from 0.01 % to 0.1 % after surface passivation of the thin film with EDAI (Figure 4a), which corresponds to a large reduction of non-radiative losses. We note that PLQE measurements are taken on samples deposited on MeO-2PACz/glass to ensure the perovskite formation is relevant to devices, and that SEM images do not show obvious surface roughening which could otherwise promote better light outcoupling (Supplementary Fig. 11). We then thermally evaporate $C_{60}$ on top of the perovskite to have a complete device stack and observe the PLQE drops to 0.02 %. On the contrary, the PLQE of the device stack without EDAI treatment is below our detection limit (<< 0.005%), proving the passivation effect of EDAI (Supplementary Fig. 12). This result is consistent with previous reports which reveal that interfacial losses between perovskite and C60 are severe if unmitigated.[44] XRD measurements show some incorporation of iodide into the perovskite upon EDAI passivation (Supplementary Fig. 13),[45] but no low angle peak is observed which excludes significant formation of 2D perovskite on the surface as reported by others.[46] Finally, TRPL measurements show prolonged charge carrier lifetimes in the EDAI-passivated sample (Supplementary Fig. 14), strengthening the viability of the approach to increase charge carrier diffusion lengths in eventual devices under operation.

We fabricate solar cells where the 1.77 eV evaporated perovskite is passivated with EDAI and show the JV curves in Figure 4b. We observe a substantial improvement in $V_{OC}$ and Fill Factor (FF) with respect to the unpassivated sample. In particular, the $V_{OC}$ reaches 1.26 V, which is the highest value reported so far to the best of our knowledge in a p-i-n, MA-free perovskite solar cell processed by vapour deposition (Supplementary Fig. 15). Figure 4c shows a comparison between the Quasi-Fermi level splitting value extracted from PLQE data (see Methods for details on the calculations) and the actual device $V_{OC}$ for both the control and the passivated sample. A difference between those values relates to the relative importance of intrinsic perovskite non-radiative recombination processes and the effect of the additional interface introduced by the $C_{60}$ layer with energetic offsets.[47] Interestingly, EDAI-treated devices gain 40 mV with respect to the control device as a result of surface passivation, following the trend observed in PL. Further analysis of the EQE curves shows a concomitant reduction in Urbach energy from 17.0 to 15.5 meV (Supplementary Fig. 16). We note that the perovskite bandgap slightly reduces from 1.77 eV to 1.76 eV as a result of iodine incorporation upon EDAI passivation (Supplementary Fig. 17), consistent with our XRD results.

To better understanding the passivation effect on the FF, we conduct a light intensity dependent measurement of the $V_{OC}$ and extract the ideality factor (Figure 4d). Using these data to extract pseudo-JV curves (see methods and Supplementary Fig. 18),[48] we deconvolute the effect of charge transport and non-radiative losses within the devices, showing 79.0 % and 85.5% FF when there is no charge transport loss. Figure 4e summarises the results, highlighting the reduction in the non-radiative losses for the EDAI-passivated samples. In contrast, we have identified an absolute reduction in FF by 10% from charge transport losses in both control and EDAI passivated samples, indicating that further optimisation to find ideal contact layers is still required. The EDAI passivation strategy also better stabilises the device $V_{OC}$ under light soaking compared to the unpassivated control (Figure 4f). We observe a 1.4% absolute increase in average PCE for the EDAI-passivated solar cells compared to the control when comparing batch-to-batch variation (Supplementary Fig. 19), demonstrating the reproducibility of EDAI passivation.

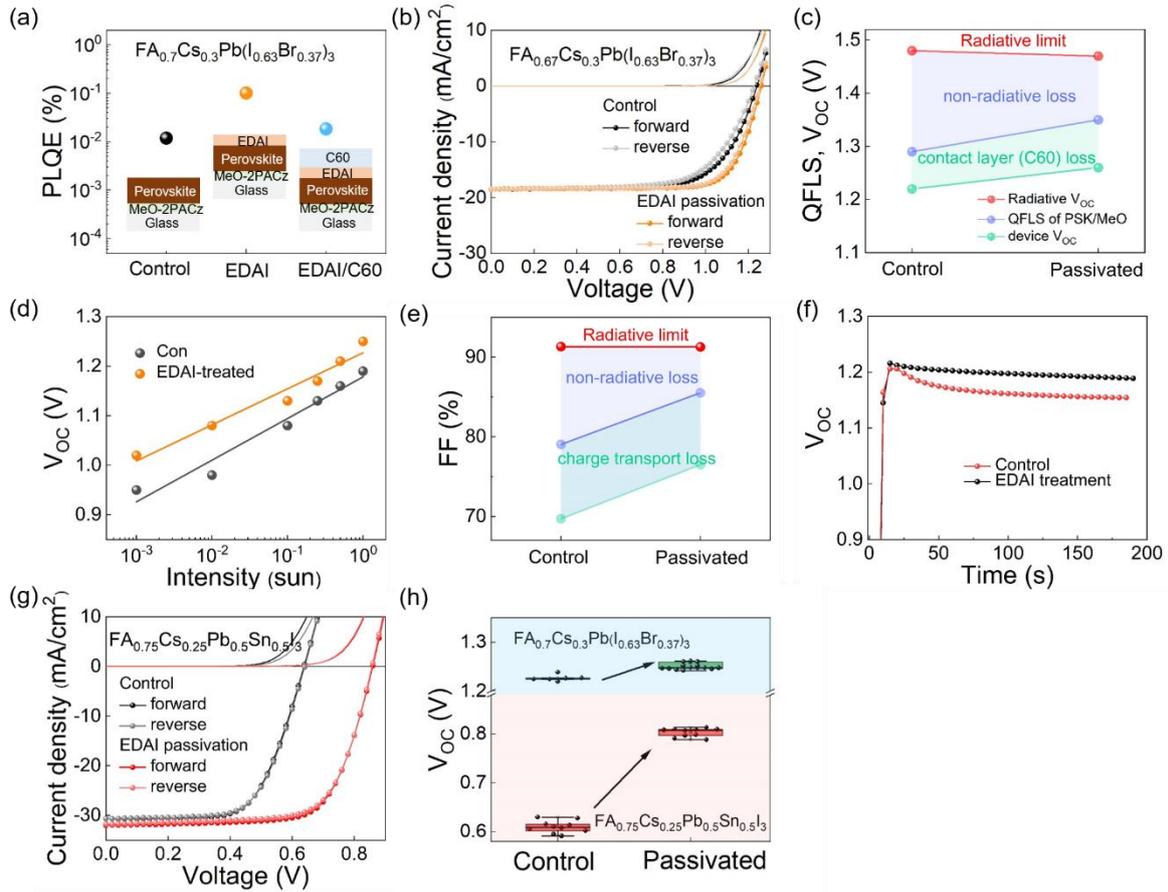

Figure 4. (a) PLQE values for $FA_{0.7}Cs_{0.3}Pb(I_{0.64}Br_{0.36})_3$ (1.77 eV) evaporated perovskite film, with EDAI treatment and with EDAI treatment plus C60 contact layer. JV parameters are shown in Table 2. (b) J-V curves for control and EDAI-passivated $FA_{0.7}Cs_{0.3}Pb(I_{0.64}Br_{0.36})_3$ based solar cell in the dark (no symbols) and under 1 sun AM 1.5 G illumination (line and symbols). See device parameters in Table 2. Radiative limit, pseudo (non-radiative loss) and experimental values of the Voc (QFLS) (c) and FF (e) for control and EDAI-treated devices. (d) Voc as a function of incident light intensity, showing an ideal factor of 1.46 and 1.27, and (f) as a function of time under AM 1.5 G illumination for control and EDAI-treated devices. (g) J-V curves for control and EDAI-passivated, $FA_{0.75}Cs_{0.25}Pb_{0.5}Sn_{0.5}I_3$ solar cell in the dark (no symbols) and under AM 1.5 G illumination (line and symbols). See device parameters in Table 2. (h) $V_{OC}$ statistics for devices based on $FA_{0.7}Cs_{0.3}Pb(I_{0.64}Br_{0.36})_3$ and $FA_{0.75}Cs_{0.25}Pb_{0.5}Sn_{0.5}I_3$ with and without EDAI passivation.

**Device performance of all-perovskite tandem solar cells**

In order to demonstrate a narrow bandgap subcell suitable for a tandem configuration, we first develop devices based on an ITO/2-PACz/FA$_{0.75}$Cs$_{0.25}$Pb$_{0.5}$Sn$_{0.5}$I$_3$/C60/BCP/Cu architecture where the perovskite in this case is deposited by solution processing. For the perovskite, we observe grain size around 500 nm, PL emission at 970 nm and bandgap at 1.28 eV (Supplementary Fig. 20). The EDAI passivation substantially improves the Voc in these narrow bandgap perovskite solar cells, increasing the champion PCE from 12.6 % to 19.4% (Figure 4g and Supplementary Fig. 19) and average from 11.2% to 18.4% (Figure 4h).[32] We note that we do not see any morphology variation or EQE onset shift after EDAI passivation (Supplementary Fig. 22 and 23). We see a reduction in VOC loss from 382 mV to 140 mV and Urbach energy from 21.5 meV to 20 meV (Supplementary Fig. 21) of the perovskite absorber when comparing devices without and with EDAI passivation, respectively, consistent with reduced non-radiative loss and electronic disorder in the perovskite films after EDAI treatment.

Table 2. Champion PV performance metrics for evaporated wide bandgap (1.77 eV) and solution processed narrow bandgap (1.28 eV) perovskite solar cells with and without EDAI treatment, and 2-terminal all-perovskite tandem solar cells.

| | \multicolumn{5}{c}{Wide bandgap (1.77 eV/1.76 eV) perovskite solar cells} | | | | |
|---|---|---|---|---|---|
| | $V_{OC}$ (V) | $J_{SC}$ (mA/cm$^2$) | FF (%) | PCE (%) | $V_{OC}$ loss (mV) |
| Control | 1.24 | -18.5 | 69.7 | 16.0 | 230 |
| EDAI-treated | 1.26 | -18.5 | 76.5 | 17.8 | 190 |
| | \multicolumn{5}{c}{Narrow bandgap (1.28 eV) perovskite solar cells} | | | | |
| | $V_{OC}$ (V) | $J_{SC}$ (mA/cm$^2$) | FF (%) | PCE (%) | $V_{OC}$ loss (mV) |
| Control | 0.64 | -30.6 | 64.3 | 12.6 | 380 |
| EDAI-treated | 0.86 | -32.0 | 70.6 | 19.4 | 140 |
| | \multicolumn{5}{c}{All-perovskite tandem solar cell} | | | | |
| | $V_{OC}$ (V) | $J_{SC}$ (mA/cm$^2$) | FF (%) | PCE (%) | |
| | 2.06 | -15.2 | 76.9 | 24.1 | |

We build a two-terminal monolithic all-perovskite tandem solar cells based on our optimised evaporated wide gap (1.77 eV) and solution processed narrow bandgap (1.28 eV) perovskite subcells, with a SnO$_x$ interconnection layer deposited by atomic layer deposition (ALD).[49,50] We note that the ALD-SnO$_x$ does not affect the performance of the wide bandgap subcell (Supplementary Fig 25). The architecture of the tandem device is shown in Figure 5a and is comprised of ITO/MeO-2PACz/FA$_{0.7}$Cs$_{0.3}$Pb(I$_{0.64}$Br$_{0.36}$)$_3$/EDAI/C60/ALD-SnO$_x$/Au/PEDOT:PSS/ FA$_{0.75}$Cs$_{0.25}$Pb$_{0.5}$Sn$_{0.5}$I$_3$/C60/BCP/Cu. A ~1 nm Au cluster layer is used between the SnO$_x$ and the PEDOT:PSS layers to improve charge recombination and enhance device V$_{OC}$ and FF.[49] We also note that the choice of Au (as opposed to Cu, for instance) for the recombination junction is important to ensure good charge transport (Supplementary Fig. 26 and Table S3). We found that employing 2-PACz for the narrow bandgap subcell in a tandem architecture has a negative impact on charge transport, which we attribute to phosphonic acid groups not anchoring well to the Au clusters.[51] Therefore, we utilise PEDOT:PSS instead of the 2-PACz. Figure 5b displays an SEM cross section image of the tandem device under study, where the thickness of the wide bandgap and narrow bandgap perovskites are 300 nm and 800 nm, respectively, to match the current of each subcell. Figure 5c shows the J-V curves of the champion all-perovskite tandem solar cell, showing a PCE of 24.1%, with an excellent $V_{OC}$ of 2.06 V, a $J_{SC}$ of 15.2 mA/cm$^2$ and a FF of 76.9 % from forward scan direction and negligible hysteresis between the forward and backward scans. The integrated JSC extracted for the wide bandgap and narrow bandgap subcells from the EQE spectra are 14.5 and 14.9 mA/cm$^2$, respectively (Figure 5d). This PCE is the highest reported value so far for an all-perovskite tandem solar cell where at least one subcell is prepared by vacuum deposition. Comparing to the sum Voc of the champion subcells, the tandem device shows a small voltage loss of 60 mV, which we attribute to the interconnecting layer that requires further optimisation. We observe a stabilised performance output PCE of 23.2% at a fixed bias of 1.74 V (Supplementary Fig. 27)

This result constitutes the first and highest-performing integration of an evaporated building block in an all-perovskite tandem architecture where the bandgaps of the absorbers employed harvest complementary regions of the spectrum. The high optical quality, pin-hole free character, and finely controlled thickness (cf. Figure 5b) of the evaporated subcell enables the subsequent deposition of recombination layers and narrow bandgap subcells to attain efficient all-perovskite tandem solar cells. The device statistic data for the PCE and the VOC, and JSC and FF across four batches are displayed in Figure 5e and 5f, showing a standard deviation of 1.5 % in PCE thus confirming reasonable batch-to-batch reproducibility.

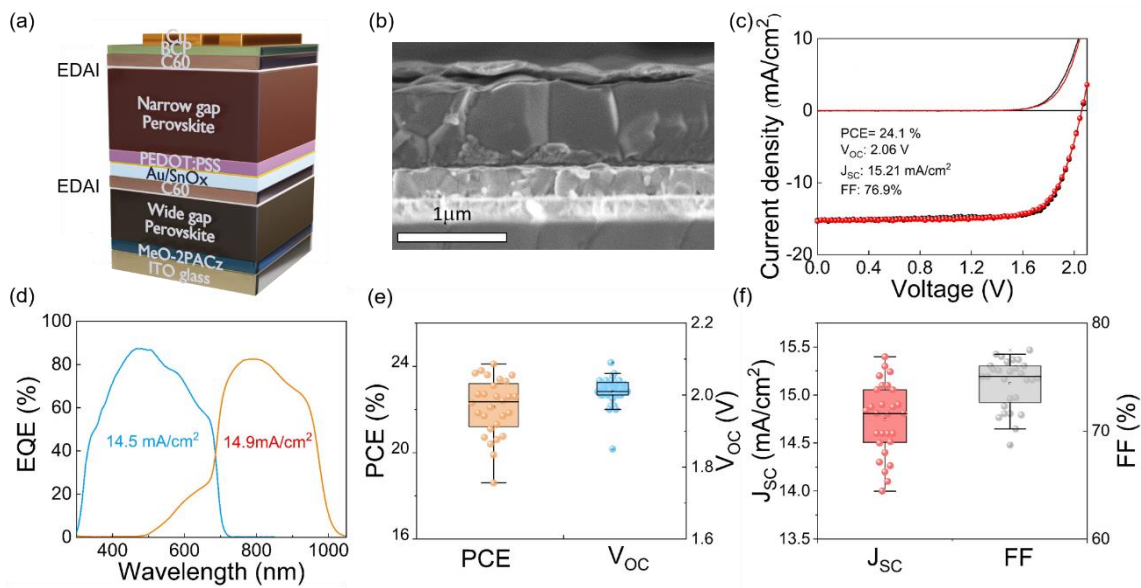

Figure 5. (a) Schematics and (b) cross-sectional SEM image displaying the architecture of the all-perovskite tandem solar cell. (c) J-V curve of the champion tandem device, reaching a 24.1% PCE for forward scan. (d) EQE spectra of the evaporated based wide bandgap (blue curve) and solution based narrow bandgap (orange) perovskite subcells comprising the all-perovskite tandem device. The distribution of device performance across four batches (28 devices) is shown in Figure 5(e) and 5(f) with an average PCE of 22.1%.

**Conclusion**

Our work employs a dual-interface treatment to maximise the performance of evaporated perovskite devices, which show great promise when implemented into tandem architectures. We employ a 4-source vacuum deposition method to demonstrate $FA_{0.7}Cs_{0.3}Pb(I_xBr_{1-x})_3$ perovskites of tunable bandgap. Engineering the device architecture via use of a MeO-2PACz layer as HTM demonstrates a 20.7% PCE in a 1.62 eV bandgap perovskite solar cell, which is highest MA-free device performance in a multi-source evaporated system. Several evaporation sources enable fine tuning of the halide content and we use it to report a phase stable $FA_{0.7}Cs_{0.3}Pb(I_{0.64}Br_{0.36})_3$ with a 1.77 eV bandgap and minimised non-radiative losses when treated with EDAI. This passivation method is versatile and reproducible, and we extend it to Pb/Sn based narrow bandgap perovskite solar cells to build a 2-terminal tandem solar cell that shows a PCE of 24.1% with an excellent $V_{OC}$ of up to 2.06 V. Our result is a key step towards all-vapour deposited tandems and encourages future work to develop narrow bandgap perovskites benefiting from the scalable, conformal and reproducible character of vacuum deposition methods. These systems open a myriad of possibilities for enhanced modularity including exploring new recombination layers not compatible with solution processed perovskites, and integration of advanced photonic strategies to push perovskite photovoltaics to their limits.

**Acknowledgements**

Y.-H.C. acknowledges the Taiwan Cambridge Trust and Rank Prize fund. K.F. acknowledges a George and Lilian Schiff Studentship, Winton Studentship, the Engineering and Physical Sciences Research Council (EPSRC) studentship, Cambridge Trust Scholarship and Robert Gardiner Scholarship. M.A. acknowledges funding from the Marie Skłodowska-Curie actions 27 (grant agreement no. 841386) under the European Union's Horizon 2020 research and innovation programme, and support by the Royal Academy of Engineering under the Research Fellowship programme. Part of this work was undertaken using equipment facilities provided by the Henry Royce Institute, via the grant Henry Royce Institute, Cambridge Equipment: EP/P024947/1 with additional funding from the "Centre for Advanced Materials for Integrated Energy Systems (CAM-IES)" (EP/P007767/1). We thanks Steve Hawes for technical support. S.D.S. acknowledges the Royal Society and Tata Group (grant no. UF150033). The work has received funding from the European Research Council under the European Union's Horizon 2020 research and innovation programme (HYPERION, grant agreement no. 756962). The authors acknowledge the EPSRC (EP/R023980/1, EP/S030638/1, EP/T02030X/1) for funding. We acknowledge the Diamond Light Source (Didcot, Oxfordshire, UK) for providing beamtime at the I14 Hard X-ray Nanoprobe facility through proposals sp20420 and mg28521. We would like to thank the beamline scientists Dr Julia E. Parker, Dr Paul D. Quinn and Dr Jessica M. Walker for their support with nXRF measurements. For the purpose of open access, the author has applied a Creative Commons Attribution (CC BY) license to any Author Accepted Manuscript version arising from this submission

**Contributions**

Y.-H.C., M.A. and S.D.S conceived and designed the work. Y.-H.C. prepared and optimised evaporated perovskite films, solution-processed perovskite films, contacts and transport layers including those deposited by ALD. Y.-H.C. designed, fabricated and optimised the wide bandgap, narrow bandgap and tandem device architectures, and conducted the device characterisation. Y.-H.C collected, analysed and interpreted SEM, XRD and PL(QE) data. M.A. and K.F. collected, analysed and interpreted n-XRF data. Y.-H.C and H.S. collected, analysed and interpreted TRPL data. H.S. and A.A. set up the TRPL equipment and performed calibrations. B. R. provided input in the ALD process. Y.-H.C, M.A. and S.D.S wrote the manuscript with comments from all the authors.

**Competing interest declaration**

S.D.S is a co-founder of Swift Solar, Inc.

**Data availability**

The data and code that support the findings of this study are available at [DOI link to be inserted at publication] in the University of Cambridge Apollo repository.